\documentclass[pra,twocolumn,superscriptaddress,aps,showpacs]{revtex4-1}

\usepackage{hyperref}
\usepackage{graphicx}
\usepackage{amsmath}
\usepackage{amsfonts}
\usepackage{amssymb}
\usepackage{wasysym}
\usepackage{times}
\usepackage{lineno}
\usepackage{longtable}
\usepackage{subfigure}
\usepackage[usenames,dvipsnames]{color}
\usepackage{bm}
\everymath{\displaystyle}
\begin{document}
\title{Quantum phases of tilted dipolar bosons in 
two-dimensional optical lattice}

\author{Soumik Bandyopadhyay}
\affiliation{Physical Research Laboratory,
             Ahmedabad - 380009, Gujarat,
             India}
\affiliation{Indian Institute of Technology Gandhinagar,
             Palaj, Gandhinagar - 382355, Gujarat,
             India}
\author{Rukmani Bai}
\affiliation{Physical Research Laboratory,
             Ahmedabad - 380009, Gujarat,
             India}
\author{Sukla Pal}
\affiliation{Physical Research Laboratory,
             Ahmedabad - 380009, Gujarat,
             India}
\affiliation{Department of Physics, Centre for Quantum Science, and Dodd-Walls
             Centre for \\ Photonic and Quantum Technologies, 
             University of Otago, Dunedin 9016, New Zealand}
\author{K. Suthar}
\affiliation{Physical Research Laboratory,
             Ahmedabad - 380009, Gujarat,
             India}
\affiliation{Instytut Fizyki imienia Mariana Smoluchowskiego, \\
             Uniwersytet Jagiello\'nski, ulica \L{}ojasiewicza 11,
             30-348 Krak\'ow, Poland}
\author{Rejish Nath}
\affiliation{Indian Institute of Science Education and Research, 
             Pune - 411008, India}
\author{D. Angom}
\affiliation{Physical Research Laboratory,
             Ahmedabad - 380009, Gujarat,
             India}
\date{\today}

\begin{abstract}
 We consider a minimal model to describe the quantum phases of ultracold 
dipolar bosons in two-dimensional (2D) square optical lattices. The model is 
a variation of the extended Bose-Hubbard model and apt to study the quantum 
phases arising from the variation in the tilt angle $\theta$ of the dipolar 
bosons. At low tilt angles $0^{\circ}\leqslant\theta\apprle25^{\circ}$, the 
ground state of the system are phases with checkerboard order, which could be 
either checkerboard supersolid or checkerboard density wave. For high tilt 
angles $55^{\circ}\apprge\theta\apprge35^{\circ}$, phases with striped order of
supersolid or density wave are preferred. In the intermediate domain 
$25^{\circ}\apprle\theta\apprle35^{\circ}$ an emulsion or SF phase intervenes 
the transition between the checkerboard and striped phases. The attractive 
interaction dominates for $\theta\apprge55^{\circ}$, which renders the system 
unstable and there is a density collapse. For our studies we use Gutzwiller 
mean-field theory to obtain the quantum phases and the phase boundaries. In 
addition, we calculate the phase boundaries between an incompressible and a 
compressible phase of the system by considering second order perturbation 
analysis of the mean-field theory. The analytical results, where applicable, 
are in excellent agreement with the numerical results.
\end{abstract}

\maketitle


\section{Introduction}\label{Introduction}
 In the strongly interacting regime, neutral bosons with short range 
interactions in optical lattices exhibit two quantum phases: Mott-insulator 
(MI) and superfluid (SF)~\cite{fisher_89,jaksch_98,greiner_02_1,greiner_02_2}. 
A prototypical model, which describes the properties of such systems is the
Bose-Hubbard model (BHM)~\cite{fisher_89,jaksch_98,hubbard_63}. The model
considers nearest neighbour hopping and onsite interaction between the bosons. 
The model is, however, not suitable to describe quantum phases which have 
offsite density-density correlations, such as, density wave (DW), 
supersolid (SS) etc~\cite{penrose_56,batrouni_00, kim_04_1,kim_04_2,goral_02,
boninsegni_05,yi_07,danshita_09}. The emergence of these quantum phases and 
their stabilization require long range interactions. The interaction could be
dipole-dipole 
interaction~\cite{goral_02,yi_07, danshita_09,lahaye_09,baranov_12}, 
fermions mediated boson-boson interaction in Bose-Fermi 
mixtures~\cite{buchler_03}, etc. The former is realized in dipolar atoms
like Cr~\cite{griesmaier_05,stuhler_05,lahaye_07}, Dy~\cite{lu_11,tang_15}, 
Er~\cite{aikawa_12,baier_16}, and polar 
molecules~\cite{ospelkaus_06,danzl_08,ni_08,ospelkaus_09,chotia_12,frisch_15}.
Apart from quantum phases in optical lattices, dipolar bosons specifically 
polar molecules, offer fast and robust schemes for quantum computation
~\cite{lincoln_09,gorshkov_11,hazzard_14}. In addition, the long range and 
anisotropic nature of the dipole-dipole interaction can induce exotic magnetic 
orders. Thus, these systems are promising simulators for quantum 
magnetism~\cite{pu_01, micheli_06, barnett_06, paz_13}.
 
 The BHM with the nearest neighbour (NN) lattice sites inter-particle 
interaction and its variations are referred to as the extended Bose-Hubbard 
model (eBHM)~\cite{mazzarella_06,dutta_15}. It is a minimal model which 
harbours phases with off site density-density correlations. Based on this 
model several theoretical studies have analyzed the equilibrium phases of 
bosons in optical lattices and their stability 
properties~\cite{sengupta_05,scarola_05,kovrizhin_05,scarola_06,menotti_07,
iskin_11,trefzger_11}, and dynamics of the quantum phase transitions by
quenching system parameters~\cite{shimizu_18_1,shimizu_18_2}. 
In 2D this is equivalent to dipole-dipole 
interaction limited to the NN interaction and with the dipoles aligned 
perpendicular to the lattice plane. And, such systems exhibit checkerboard 
order in the DW and SS phases. Thus, a minimal model to describe quantum 
phases of dipolar bosons in optical lattices is to limit the interaction to 
NN. This is the system we consider in our present work. In previous studies, 
the quantum phases of lattice bosons with anisotropic dipolar interaction and 
their stability has been analyzed~\cite{goral_02,yi_07,danshita_09}. In 
addition, the phase diagrams for the dipolar bosons in 2D square optical 
lattice with staggered flux in the minimal model has been 
done~\cite{tieleman_11}. A recent work~\cite{zhang_15} reported the 
equilibrium phases of the hardcore dipolar bosons at half filling in a 2D 
optical lattice with the variation of tilt angle. And, they reported DW phase
with checkerboard and stripe order.  However, the experimental observations 
are in the soft-core regime~\cite{baier_16}. In this experiment Baier 
et al.~\cite{baier_16} have realized the eBHM for the strongly magnetic Er 
atoms in a 3D optical lattice and observed NN interaction as a genuine 
consequence of the long-range dipolar interactions. And, they also vary tilt 
angle of the dipolar atoms to examine the effect of anisotropic dipole-dipole 
interaction on the SF-MI phase transition. 

  Motivated by the experimental realization, we investigate the quantum phases 
of tilted softcore dipolar bosons in a 2D square optical lattice. Hence, our 
work addresses a key research gap in the physics of softcore dipolar bosons in 
the strongly interacting domain. We show that the system exhibits compressible 
checkerboard SS (CBSS) and striped SS (SSS) phases in addition to the 
incompressible checkerboard DW (CBDW) and striped DW (SDW) phases. Our results 
can be experimentally examined since tilting the dipoles have become a standard
tool box to understand physics of ultracold dipolar bosons and 
fermions~\cite{bismut_12,aikawa_14,velji_18}.
 
 We have organized the remainder of this article as follows. In 
Sec.~\ref{sec_model} we discuss the zero-temperature Hamiltonian of the
minimal model. The Sec.~\ref{sec_theory} provides a brief account of the 
Gutzwiller mean-field theory, and the quantum phases of the model. Then, in 
the later part of the section, we discuss the mean-field decoupling theory
to calculate the compressible-incompressible phase boundaries analytically.
The Sec.~\ref{sec_numeric} describes the numerical procedures adopted to 
solve the model. The phase diagrams and key results of our work are 
discussed in Sec.~\ref{sec_results}. We, then, conclude in 
Sec.~\ref{conclusions}. 

\begin{figure}[ht]
    \includegraphics[height = 4.5cm]{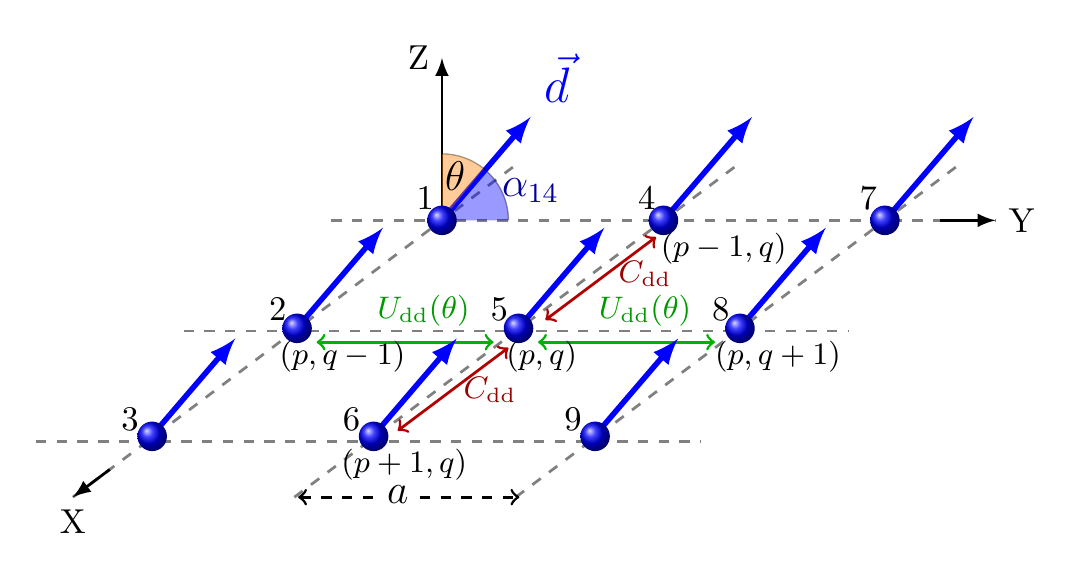}
    \caption{(Color online)
             Schematics of the dipolar bosons in two dimensional square optical 
             lattice with dipolar interaction among the bosons at 
             nearest-neighbour (NN) lattice sites. We consider the dipoles are 
             polarized in the y-z plane and the angle subtended by the 
             direction of the dipole moments (polarization axis) with the 
             z-axis is the tilt angle $\theta$. The tilt angle is illustrated 
             by the orange colored shaded sector. The angle between the 
             polarization axis and the vector $(\vec{r}_{4}-\vec{r}_{1})$, 
             $\alpha_{14}$, is marked by the blue colored shaded sector. The 
             dipolar interaction between the bosons at lattice sites $(p,q)$ 
             and $(p\pm1, q)$ is $C_{\rm dd}$, whereas the interaction between 
             the bosons at lattice sites $(p,q)$ and $(p, q\pm1)$ is 
             $U_{\rm dd}(\theta) = C_{\rm dd}(1-3\sin^{2}\theta)$}
    \label{schm_dipolar}
\end{figure}


\section{Theoretical Model}\label{sec_model}
We consider charge neutral, polarized dipolar bosons loaded in a 2D
square optical lattice with lattice constant $a$. At zero temperature, the 
physics of such a system is well described by the lowest band Bose-Hubbard 
model (BHM) with dipolar interaction. The grand canonical Hamiltonian of the
system is~\cite{fisher_89,goral_02,boninsegni_05,yi_07,danshita_09}:
\begin{equation}
   \hat{H} = -J\sum_{\langle ij\rangle}
                 (\hat{b}_{i}^{\dagger}\hat{b}_{j} + {\rm H.c.})
                 - \sum_{i} \mu\hat{n}_{i} + \hat{H}_I,
 \label{bhm_hamil}                
\end{equation}
where $i\equiv(p,q)$ and $j\equiv(p',q')$ denote the lattice indices,
$\hat{b}_{i}$ ($\hat{b}_{i}^{\dagger}$) and $\hat{n}_{i}$ are bosonic 
annihilation (creation) and occupation number operators, and 
$\langle...\rangle$ denotes sum over NN lattice sites. In addition, $J$ and 
$\mu$ are the strength of the hopping and chemical potential, respectively.
The last term is the interatomic interaction Hamiltonian
\begin{equation}
  \hat{H}_I = \sum_{i}\frac{U}{2}\hat{n}_{i}(\hat{n}_{i}-1)
              + \frac{C_{\rm dd}}{2}\sum_{ij} \hat{n}_{i}\hat{n}_{j}
                 \frac{(1-3{\rm cos}^2\alpha_{ij})}
                 {|\vec{r}_{j}-\vec{r}_{i}|^{3}},
 \label{h_int}                
\end{equation}
where, $U$ and $C_{\rm dd} \propto d^{2}/a^{3}$ are the strengths of the onsite 
and dipolar interactions, respectively.  Here,  $d$ is the magnitude of the
induced dipole moment, and $\alpha_{ij}$ is the angle between the polarization 
axis and the vector $(\vec{r}_{j}-\vec{r}_{i})$. In units of $a$ the position
vectors of the lattices $\vec{r}_{i} \equiv (p\hat{e}_x + q\hat{e}_y)$ 
and $\vec{r}_{j} \equiv (p'\hat{e}_x + q'\hat{e}_y)$.

In our study, for simplicity, we limit the dipolar interaction to NN sites.
Then,
\begin{equation}
   \hat{H}_I = \sum_{i}\frac{U}{2}\hat{n}_{i}(\hat{n}_{i}-1)
               + \frac{C_{\rm dd}}{2}\sum_{\langle ij\rangle}
                 \hat{n}_{i}\hat{n}_{j}
                 (1-3{\rm cos}^2\alpha_{ij}).
 \label{bhm_nn}                
\end{equation}
This minimal model is apt for studying the quantum phases of dipolar bosons
emerging from the anisotropic nature of the dipolar interaction. In addition, 
we consider the dipoles are polarized in the $yz$-plane as illustrated in 
Fig~(\ref{schm_dipolar}), and define the angle between the $z$-axis and 
polarization axis as the tilt angle $\theta$. With this choice, $\alpha_{ij}$ 
changes as a function of $\theta$, which can be varied by changing the 
orientation of the applied magnetic field. Then, the NN interaction along 
$x$-axis is always repulsive, constant, and independent of $\theta$. Whereas, 
along the $y$-axis the NN interaction is 
$U_{\rm dd}(\theta) = C_{\rm dd}(1 - 3\sin^2\theta)$. It varies from 
$C_{\rm dd}$ to $-2C_{\rm dd}$ as $\theta$ is tuned from $0^{\circ}$ to 
$90^{\circ}$. And, the zero of $U_{\rm dd}(\theta)$ occurs when 
$\theta = \theta_{\rm M} = \sin^{-1}\big(1/\sqrt{3}\big) 
\approx 35.3^{\circ}$.
This angle is referred to as the magic angle~\cite{ueda_2010} and at this 
tilt angle the interaction arising from dipolar interaction is absent along
the $y$-axis. Thus, the interaction along $y$-axis is repulsive when 
$\theta<35.3^{\circ}$, and attractive for $\theta>35.3^{\circ}$.


\section{Theoretical methods}\label{sec_theory}
\subsection{Gutzwiller mean-field theory} 

To solve the model, we consider site decoupled mean-field (MF) 
approximation~\cite{fisher_89,rokhsar_91,sheshadri_93,bai_18,pal_19,suthar_19}.
For this, the bosonic annihilation operator of site $(p,q)$, $\hat{b}_{p,q}$, 
is decomposed to a mean-field $\phi_{p,q}$ and fluctuation operator 
$\delta \hat{b}_{p,q}$ as 
$\hat{b}_{p,q}=\langle \hat{b}_{p,q}\rangle + \delta \hat{b}_{p,q} = 
\phi_{p,q} + \delta \hat{b}_{p,q}$. A similar decomposition is applied to 
$\hat{b}^{\dagger}_{p,q}$ and $\hat{n}_{p,q}$. It is to be mentioned that, 
here after we adopt the explicit notation $(p,q)$ to denote a lattice site in 
2D. To obtain the MF Hamiltonian, we use the decomposed operators in $\hat{H}$ 
and neglect the terms which are quadratic in fluctuation operators. 
Then, the MF Hamiltonian of the system is
\begin{widetext}
 \begin{eqnarray}
  \hat{H}_{\rm MF} 
            &=& \sum_{p,q}\Big\{
              -J\left[\left(\hat{b}_{p+1,q}^{\dagger}\phi_{p,q}
              +\phi^{*}_{p+1,q}\hat{b}_{p,q} - \phi^{*}_{p+1,q}\phi_{p,q}\right)
              + \left(\hat{b}_{p,q+1}^{\dagger}\phi_{p,q}
              +\phi^{*}_{p,q+1}\hat{b}_{p,q} - \phi^{*}_{p,q+1}\phi_{p,q}\right)
              + {\rm H.c.}\right]
              - \mu\hat{n}_{p,q}
              \nonumber\\
            &+& \frac{U}{2}\hat{n}_{p,q}(\hat{n}_{p,q}-1)
              + \frac{C_{\rm dd}}{2}\Big[
              \Big(\hat{n}_{p+1,q}\langle\hat{n}_{p,q}\rangle
              + \langle\hat{n}_{p+1,q}\rangle\hat{n}_{p,q}
              -\langle\hat{n}_{p+1,q}\rangle
              \langle\hat{n}_{p,q}\rangle\Big)
              +\Big(\hat{n}_{p-1,q}\langle\hat{n}_{p,q}\rangle
              + \langle\hat{n}_{p-1,q}\rangle\hat{n}_{p,q}
              \nonumber\\
            &-&\langle\hat{n}_{p-1,q}\rangle
              \langle\hat{n}_{p,q}\rangle\Big)\Big]
              +\frac{U_{\rm dd}(\theta)}{2}\Big[
              \Big(\hat{n}_{p,q+1}\langle\hat{n}_{p,q}\rangle
              +\langle\hat{n}_{p,q+1}\rangle\hat{n}_{p,q}
              -\langle\hat{n}_{p,q+1}\rangle\langle\hat{n}_{p,q}\rangle\Big)
              +\Big(\hat{n}_{p,q-1}\langle
              \hat{n}_{p,q}\rangle
              +\langle\hat{n}_{p,q-1}\rangle\hat{n}_{p,q}
              \nonumber\\ 
            &-&\langle\hat{n}_{p,q-1}\rangle\langle\hat{n}_{p,q}\rangle\Big)
              \Big]  
              \Big\}.
  \label{bhm_hamil_mf}
 \end{eqnarray}
\end{widetext}
This can be written in terms of single-site Hamiltonians as
 \begin{equation}
  \hat{H}_{\rm MF} = \sum_{p,q}\hat{h}_{p,q}, 
 \end {equation}  
where $\hat{h}_{p,q}$ is the single-site Hamiltonian of site $(p,q)$, which 
can be expressed as 

\begin{eqnarray}
  \hat{h}_{p,q} = && -J\left[\left(\phi_{p+1, q}^*\hat{b}_{p, q} 
                     + \phi_{p, q+1}^{*} \hat{b}_{p, q} 
                     \right) + {\rm H.c.}
                       \right] -\mu \hat{n}_{p, q}
                     \nonumber\\
                   &&
                     + \frac{U}{2}\hat{n}_{p, q}(\hat{n}_{p, q}- 1)
                     + \frac{C_{\rm dd}}{2}\hat{n}_{p,q}
                      \Big(\langle\hat{n}_{p+1,q}\rangle
                     \nonumber\\ 
                   &&  
                     +\langle\hat{n}_{p-1,q}\rangle\Big)
                     + \frac{U_{\rm dd}(\theta)}{2}\hat{n}_{p,q} 
                      \Big(\langle\hat{n}_{p,q+1}\rangle 
                     + \langle\hat{n}_{p,q-1}\rangle \Big),
                     \nonumber\\
  \label{bhm_hamil_explicit}
\end{eqnarray}
where we have dropped the pure MF terms. These terms shifts the ground state 
energy and play no role in determining the ground state or the phase diagrams 
of the system. We can solve the model by diagonalizing the single-site
Hamiltonians coupled through the mean-field $\phi_{p,q}$ self consistently. To 
obtain the ground state of the system, we consider site  dependent Gutzwiller 
ansatz 
\begin{equation}
  |\Psi_{\rm GW}\rangle = \prod_{p,q}|\psi_{p,q}\rangle =\prod_{p,q}
                       \sum_{n=0}^{(N_{\rm b}-1)}c^{(p,q)}_n|n\rangle_{p,q},
  \label{gw_state}
\end{equation}
where $\{|n\rangle_{p,q}\}$ are the occupation number basis states at site 
$(p,q)$, $N_{\rm b}$ is the total number of local Fock states used in the 
computation, and $c^{(p,q)}_n$ are complex coefficients of the ground state 
$|\psi_{p,q}\rangle$. The normalization of $|\Psi_{\rm GW}\rangle$ is ensured 
by considering site-wise normalization condition 
 \begin{equation}
  \langle\psi_{p,q}|\psi_{p,q}\rangle = \sum_{n=0}^{(N_{\rm b}-1)}
                                        |c^{(p,q)}_n|^{2} 
                                      = 1.
  \label{normalisation}
 \end{equation}
Then, the mean-field or superfluid order parameter $\phi_{p,q}$ and the average 
occupancy $n_{p,q}$ at the lattice site $(p,q)$ are 
\begin{eqnarray}
  \phi_{p,q} &=& \langle\Psi_{\rm GW}|\hat{b}_{p,q} |\Psi_{\rm GW}\rangle 
              = \sum_{n=1}^{(N_{\rm b}-1)}\sqrt{n} 
               {c^{(p,q)}_{n-1}}^{*}c_n^{(p,q)}, \nonumber \\
   n_{p,q}   &=& \langle\Psi_{\rm GW}| \hat{n}_{p,q}
                                |\Psi_{\rm GW}\rangle 
              = \sum_{n = 0}^{(N_{\rm b}-1)}n |c_n^{(p,q)}|^2.
  \label{av_occup}
 \end{eqnarray}
As the name indicates, $\phi_{p,q}$ is non-zero quantity in the SF phase, 
and from the definition, it is an indicator of the number fluctuation. Hence,
it is a measure of the long range phase coherence in the system. In other
words, the SF phase has off diagonal long range order (ODLRO).


\subsection{Quantum phases and their characterization}
 In absence of the dipolar interaction, depending on $J/U$ there are two 
ground state quantum phases of the system: the superfluid (SF) and 
Mott-insulator (MI) phases. The key distinction between these two phases is 
that  $\phi_{p,q}$, as mentioned earlier, is finite in SF phase. But, it is 
zero in the MI phase. In a homogeneous lattice system, density distribution of 
these two phases is uniform. However, this translational symmetry can be 
spontaneously broken with long range dipole-dipole interaction. This leads to 
the emergence of quantum phases which have periodic density modulations, such 
as, density wave (DW) and supersolid (SS). In other words, the system can 
exhibit diagonal order. Among the two phases the SS phase, in addition to the  
diagonal order, has ODLRO. Therefore, the SS phase has non-zero $\phi_{p,q}$, 
and $n_{p,q}$ has a periodic structure. On the other hand for the DW phase, 
like in the MI phase, $\phi_{p,q}$ is zero and 
$n_{p,q}$ is integer. But, unlike MI phase $n_{p,q}$ in DW show spatial 
pattern. To characterize the diagonal order in DW and SS phases, we compute 
the static structure factor
\begin{equation}
  S(\vec{k})= \frac{1}{N^2}
        \sum_{i,j}e^{{\rm i}\vec{k}.(\vec{r}_i-\vec{r}_j)}
        \langle\hat{n}_i\hat{n}_{j}\rangle,
  \label{st_fac}        
\end{equation}
where $\vec{k}\equiv (k_x, k_y)\equiv(k_x\hat{e}_x+k_y\hat{e}_y)$ is the 
reciprocal lattice vector (measured in units of $1/a$), and $N$ is the total 
number of bosons in the system. In the present study, depending on the
tilt angle $\theta$, the system has $n_{p,q}$ which is either checkerboard 
or striped. The checkerboard order breaks the translational symmetry along 
both $x$ and $y$ directions, and is characterized by a finite
value of $S(\vec{k})$ at the reciprocal lattice site $\vec{k} = (\pi, \pi)$.
In the phases having striped pattern, the translational symmetry is broken only 
along the $x$-direction. And, $S(\vec{k})$ is non-zero only for
$\vec{k} = (\pi, 0)$. Thus, the structure factors $S(\pi, \pi)$ and 
$S(\pi, 0)$ can be used to characterize the CB and striped phases. 
Like the MI phase, the DW phase is an incompressible phase of the system; 
whereas, in the SF and SS phases, the system is compressible. 
Table~(\ref{table_ch_phases}) summarizes the distinct characteristics of the 
different possible phases of the considered system.  
\begin{table}[ht]
  \begin{tabular}{l | c | c | c | c }
   \hline
   Quantum phases &${n}_{p,q}$ & $\phi_{p,q}$ & $S(\pi,\pi)$ & $S(\pi,0)$ \\
   \hline \hline
   Superfluid (SF)                 & real    & $\neq 0$ & $0$      & $0$ \\
   Mott-insulator (MI)             & integer & $0$      & $0$      & $0$ \\
   Chekerboard supersolid (CBSS)   & real    & $\neq 0$ & $\neq 0$ & $0$ \\
   Striped supersolid (SSS)        & real    & $\neq 0$ & $0$      & $\neq 0$\\
   Emulsion supersolid             & real    & $\neq 0$ & $\neq 0$ & $\neq 0$\\
   Chekerboard Density wave (CBDW) & integer & $0$      & $\neq 0$ & $0$ \\
   Striped Density wave (SDW)      & integer & $0$      & $0$      & $\neq 0$\\
   Emulsion Density wave           & integer & $0$      & $\neq 0$ & $\neq 0$\\
   \hline
  \end{tabular}
  \caption{Illustrates characteristics of different quantum phases of the
           considered system.}
  \label{table_ch_phases} 
\end{table}

\begin{figure}[ht]
    \includegraphics[height = 3.5cm]{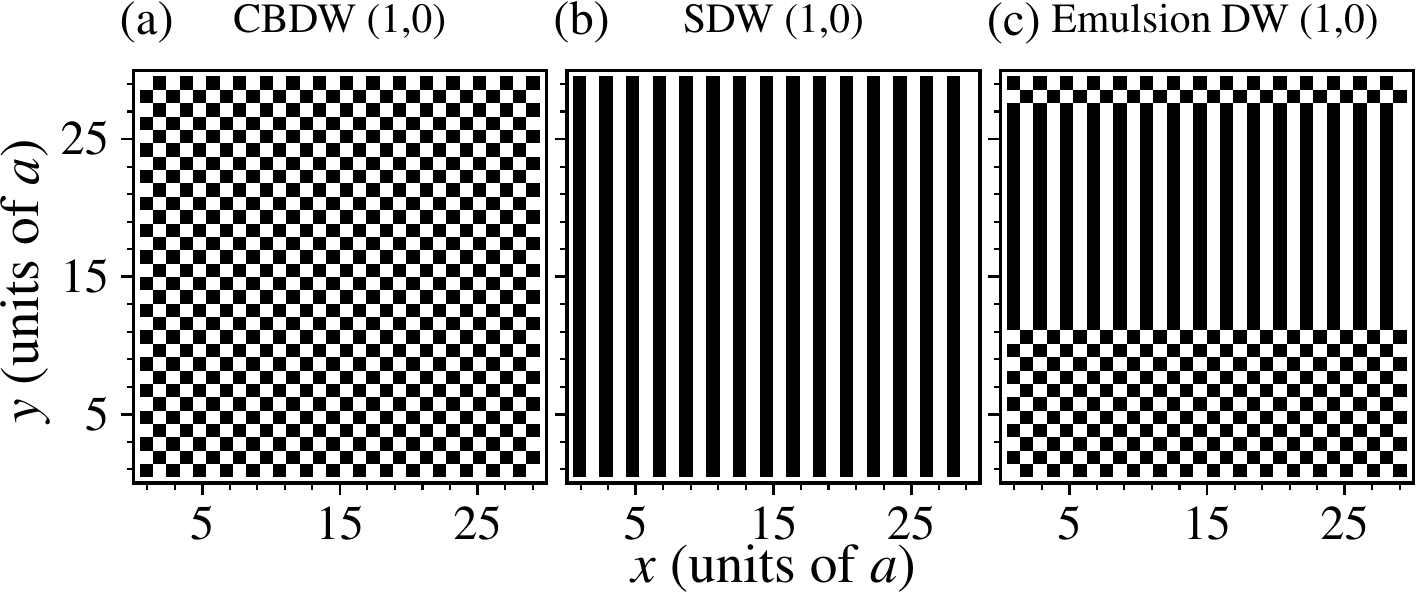}
    \caption{(Color online)
            Shows the density pattern of the system in distinct density wave 
            phases. Black squares mark those lattice sites which are vacant, 
            and white squares denote singly occupied lattice sites. The states 
            are illustrated for fixed $\mu/J=15$ and $C_{\rm dd}/U=0.8$. The 
            CBDW (1,0) and SDW (1,0) states are obtained for $J/U = 0.033$ at 
            $\theta = 0^{\circ}$ and $37^{\circ}$ respectively. The emulsion 
            phase is obtained for $J/U= 0.035$ at $\theta = 31.5^{\circ}$.
            }
    \label{den_dw_1_0}
\end{figure}
 To illustrate the density distribution in the structured phases, the density 
distribution in the CBDW (1,0), SDW (1,0) and emulsion DW (1,0) are shown in 
Fig~\ref{den_dw_1_0}. As to be expected, in Fig~\ref{den_dw_1_0}(a) the density
modulation of the CBDW (1,0) phase is along both the directions. And, in 
Fig~\ref{den_dw_1_0}(b) for the SDW (1,0) phase the density modulation is along
the $x$-axis. The emulsion phase, as shown in Fig~\ref{den_dw_1_0}(c), has 
regions with both types of density modulations. And, the simultaneous existence
of the two orders is reflected in the non-zero values of the structure factors 
$S(\pi,\pi)$ and $S(\pi,0)$. The density distribution of the checkerboard, 
striped and emulsion SS phases are also similar to the density pattern in 
Fig~\ref{den_dw_1_0}, except the densities are real number.


\subsection{Phase boundaries from mean-field decoupling theory}
 To gain additional insights on the phase transitions between compressible 
and incompressible phases we calculate the phase boundaries analytically using 
the mean-field decoupling theory~\cite{oosten_01,iskin_09_1}. A similar 
analysis can be done using other methods like strong-coupling 
expansion~\cite{freericks_96,iskin_09_1,sachdeva_12} or random phase 
approximation~\cite{iskin_09_2}. For this we use the decoupling scheme, 
described earlier,  $\hat{b}_{p,q}=\phi_{p,q} + \delta \hat{b}_{p,q}$,
$\hat{b}_{p,q}^{\dagger}=\phi_{p,q}^{*} + \delta \hat{b}_{p,q}^{\dagger}$,
and $\hat{n}_{p,q}=n_{p,q} + \delta \hat{n}_{p,q}$. Here, the  SF order 
parameter, $\phi_{p,q}$, is non-zero in SF and SS phases, but zero in the MI 
and DW phases. Then, assuming the phase transition is continuous, the phase 
boundary between a compressible ($\phi_{p,q}\neq 0$) and incompressible 
($\phi_{p,q}=0$) phase is marked by vanishing SF order parameter 
$\phi_{p,q}\to 0^{+}$. In addition, the MI and DW phases correspond to integer 
occupancies per lattice site, and are the 
exact eigenstates of the interaction and chemical potential part of
the mean-field Hamiltonian in Eq.~(\ref{bhm_hamil_mf}). Thus, the hopping 
term in the Hamiltonian can be considered as a perturbation with $\phi_{p,q}$
as the perturbation parameter. We can, then, perform a perturbative analysis
(details are given in Appendix~\ref{perturb_ana}) to obtain the order parameter
from the first order wavefunction as
\begin{equation}
  \!\phi_{p,q} = J\overline{\phi}_{p,q} \left[
              \frac{n_{p,q}+1}{U n_{p,q} - \tilde{\mu}_{p,q}}
              - \frac{n_{p,q}}{U (n_{p,q} - 1) - \tilde{\mu}_{p,q}} \right], 
  \label{phi_condition}
\end{equation}
where $\tilde{\mu}_{p,q}= \mu - V^{\rm{dip}}_{p,q}$ and
\begin{eqnarray}
  \overline{\phi}_{p,q} & = &(\phi_{p+1,q} + \phi_{p-1,q} + \phi_{p,q+1} +
  \phi_{p,q-1}), \nonumber \\
  V^{\rm{dip}}_{p,q}    & = & \frac{C_{\rm dd}}{2}(n_{p+1,q} + n_{p-1,q})
  +\frac{U_{\rm dd}(\theta)}{2}(n_{p,q+1} + n_{p,q-1}). \nonumber
\end{eqnarray}
A similar equation is obtained from the Landau procedure for continuous phase 
transition. In which case the energy functional defined as a function of 
$\phi_{p,q}$ is minimized~\cite{iskin_11,sowinski_14}. In the MI phase, the 
system has integer commensurate filling, say $n_0$, and in the SF phase it 
has uniform SF order parameter $\varphi_0$. With these considerations, 
\begin{eqnarray}
   \overline{\phi}_{p,q} &\equiv&\overline{\phi}=4\varphi_0, 
                                 \nonumber \\
   \tilde{\mu}_{p,q} &\equiv& \tilde{\mu}= 
            \mu - \left [C_{\rm dd} + U_{\rm dd}(\theta)\right ]n_0. 
                                 \nonumber
\end{eqnarray}
Since in the SF phase $\varphi_0\to0^{+}$ near the phase boundary, then from 
Eq.~(\ref{phi_condition}) the MI-SF phase boundary can be calculated from
\begin{equation}
   \frac{1}{4J} = \left[ \frac{n_0+1}{U n_0 - \tilde{\mu}}  
   - \frac{n_0}{U (n_0 - 1) - \tilde{\mu}} \right]. 
  \label{mi_sf_boundary}
\end{equation}
The solutions of the above equation defines the MI-SF boundary in the 
$\mu$-$J$ plane corresponding to the MI lobe with $n_0$ filling.
 
 To describe the phase transition from DW to SS phase, we consider two
sublattice description of the phases. That is, dipolar interaction induced 
solid order or spatially periodic modulation can be considered as if the
system has two sublattices $A$ and $B$. Each sublattice has different 
occupancies $n_A$ and $n_B$ as well as two order parameter $\varphi_{A}$ 
and $\varphi_{B}$. In the checkerboard order the periodic modulation is
along both $x$ and $y$-directions with a period of $2a$. Whereas, in the
striped order the modulation is along one of the directions. So, to obtain the
phase boundary between the SDW and SSS phases from Eq.~(\ref{phi_condition}), 
we consider striped sublattice structure. Therefore, we define 
$\overline{\phi}_{p,q} = 2(\varphi_{A} + \varphi_{B})$,
$\tilde{\mu}_A = \mu - \left [C_{\rm dd}N_{B}+ U_{\rm dd}(\theta)N_{A}\right]$
for $(p,q)\in A$ sublattice, and 
$\overline{\phi}_{p,q} =  2(\varphi_{A} + \varphi_{B})$,
$\tilde{\mu}_B = \mu - \left [C_{\rm dd}N_{A}+ U_{\rm dd}(\theta)N_{B}\right]$
for $(p,q)\in B$ sublattice. This leads to two coupled equations for 
$\varphi_{A}$ and $\varphi_{B}$:
\begin{subequations}
   \begin{equation}
    \varphi_{A} = 2(\varphi_{A} + \varphi_{B})J \left[
    \frac{n_{A}+1}{U n_{A} - \tilde{\mu}_{A}}
    - \frac{n_{A}}{U (n_{A} - 1) - \tilde{\mu}_{A}} \right],
   \label{sdw_sss_phi_a}
   \end{equation}
   \begin{equation}
    \varphi_{B} = 2(\varphi_{A} + \varphi_{B})J \left[
    \frac{n_{B}+1}{U n_{B} - \tilde{\mu}_{B}}
    - \frac{n_{B}}{U (n_{B} - 1) - \tilde{\mu}_{B}} \right].
   \label{sdw_sss_phi_b}
   \end{equation}
\end{subequations}
We solve these two equations simultaneously. In the SSS phase 
$\{\varphi_{A}, \varphi_{B}\}\to0^{+}$ across the SDW-SSS phase boundary. 
Then, the SDW-SSS phase boundary is obtained as the solution of
\begin{eqnarray}
    \frac{1}{2J} = && \left[ \frac{n_{A}+1}{U n_{A} - \tilde{\mu}_{A}}
    - \frac{n_{A}}{U (n_{A} - 1) - \tilde{\mu}_{A}} \right]
    +
    \left[ \frac{n_{B}+1}{U n_{B} - \tilde{\mu}_{B}} \right.
    \nonumber\\
    &&\left. - \frac{n_{B}}{U (n_{B} - 1) - \tilde{\mu}_{B}}
    \right].
   \label{sdw_sss_boundary}
\end{eqnarray}
Following similar reasoning, the CBDW-CBSS phase boundary is obtained as the 
solution of 
\begin{eqnarray}
  \frac{1}{16J^2} &=& \left[ \frac{n_A+1}{U n_A - \tilde{\mu}_A} 
           -\frac{n_A}{U (n_A - 1) - \tilde{\mu}_A} \right]
   \nonumber\\
   && \times \left[ \frac{n_B+1}{U n_B - \tilde{\mu}_B}
   - \frac{n_B}{U (n_B - 1) - \tilde{\mu}_B} \right].
  \label{cbdw_cbss_boundary}
\end{eqnarray}
For $\theta = 0^{\circ}$, this becomes identical to the phase boundary in
2D reported by Iskin~\cite{iskin_11}. The detailed steps of derivations to 
obtain the above equation are discussed in Appendix~(\ref{cbdw_cbss_pb}).
 
 It is to be mentioned here that close to $\theta_{\rm M}$, the system 
undergoes a checkerboard-striped transition. So, in this regime the system
can exhibit both the orders simultaneously, leading to an emulsion DW phase. 
The parameter domains of such emulsion DW phases are identified as the 
regions where Eq.~(\ref{sdw_sss_boundary}) and Eq.~(\ref{cbdw_cbss_boundary}) 
both applicable.


\section{Numerical methods}\label{sec_numeric}
 To obtain the equilibrium phase diagrams of the system, we diagonalize the
single-site Hamiltonian in 
Eq. (\ref{bhm_hamil_explicit})~\cite{bai_18,pal_19}. For this, we consider a 
guess solution of the ground state $|\Psi_{\rm GW}\rangle$ to compute the 
initial values of $\phi_{p,q}$ and $\langle\hat{n}_{p,q}\rangle$. We then 
use these values in Eq.~(\ref{bhm_hamil_explicit}), and diagonalize it to 
obtain a new ground state $|\psi_{p,q}\rangle$. Using this new state we update
$|\Psi_{\rm GW}\rangle$, and then, compute the corresponding
$\phi_{p,q}$ and $\langle\hat{n}_{p,q}\rangle$. We, then, repeat the same for
the next lattice site. This is repeated till all the lattices sites are 
covered. One such step constitutes an iteration, and the iteration is repeated
till $\phi_{p,q}$ and $\langle\hat{n}_{p,q}\rangle$ converge. Around the 
phase boundary the convergence is slow and this is remedied by considering 
larger number of iterations. To model an uniform infinite size lattice,  
we perform the above procedure on the surface of a torus by considering 
periodic boundary conditions along the $x$ and $y$-directions of the finite 
sized lattice system. In general, we have considered $12 \times 12$ 
lattice system and $N_{\rm b} = 20$ to obtain the phase diagrams. System size 
dependence of phase boundary occurs when there is an intervening emulsion phase
between two phases. For such special cases, we supplement the results from 
$12 \times 12$ lattice with the results obtained for $20\times20$ and 
$30\times30$ lattice systems.


\section{Results and discussions}\label{sec_results}
The model Hamiltonian considered has five independent parameters, namely, 
$J$, $U$, $\mu$, $C_{\rm dd}$, and $\theta$. To examine the phase diagram of 
the system in detail we scale the Hamiltonian with respect to $J$ and set 
$\mu/J = 15$. This reduces the number of independent parameters to three, 
$U/J$, $C_{\rm dd}/J$ and $\theta$. For better description, we obtain the phase 
diagrams in the $J/U$-$C_{\rm dd}/U$ plane for different values of $\theta$. 
This choice is suitable to probe the interplay between the onsite and dipolar 
interactions in determining the distinct phases of the system.


\subsection{$J/U$-$C_{\rm dd}/U$ phase diagrams} 
\begin{figure}[ht]
    \includegraphics[height = 5.8cm]{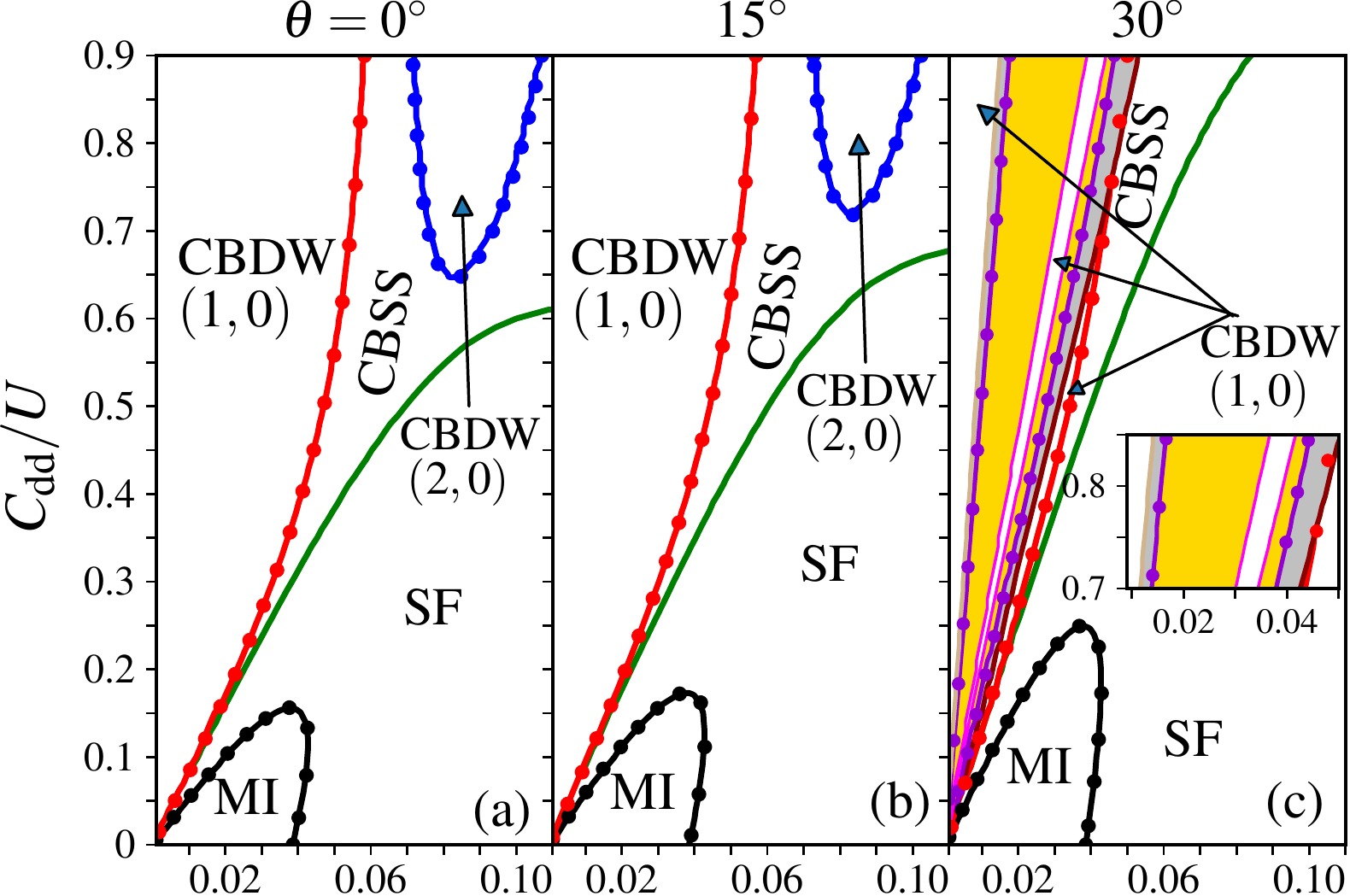}
    \vspace{0.1cm}

    \includegraphics[height = 6.3cm]{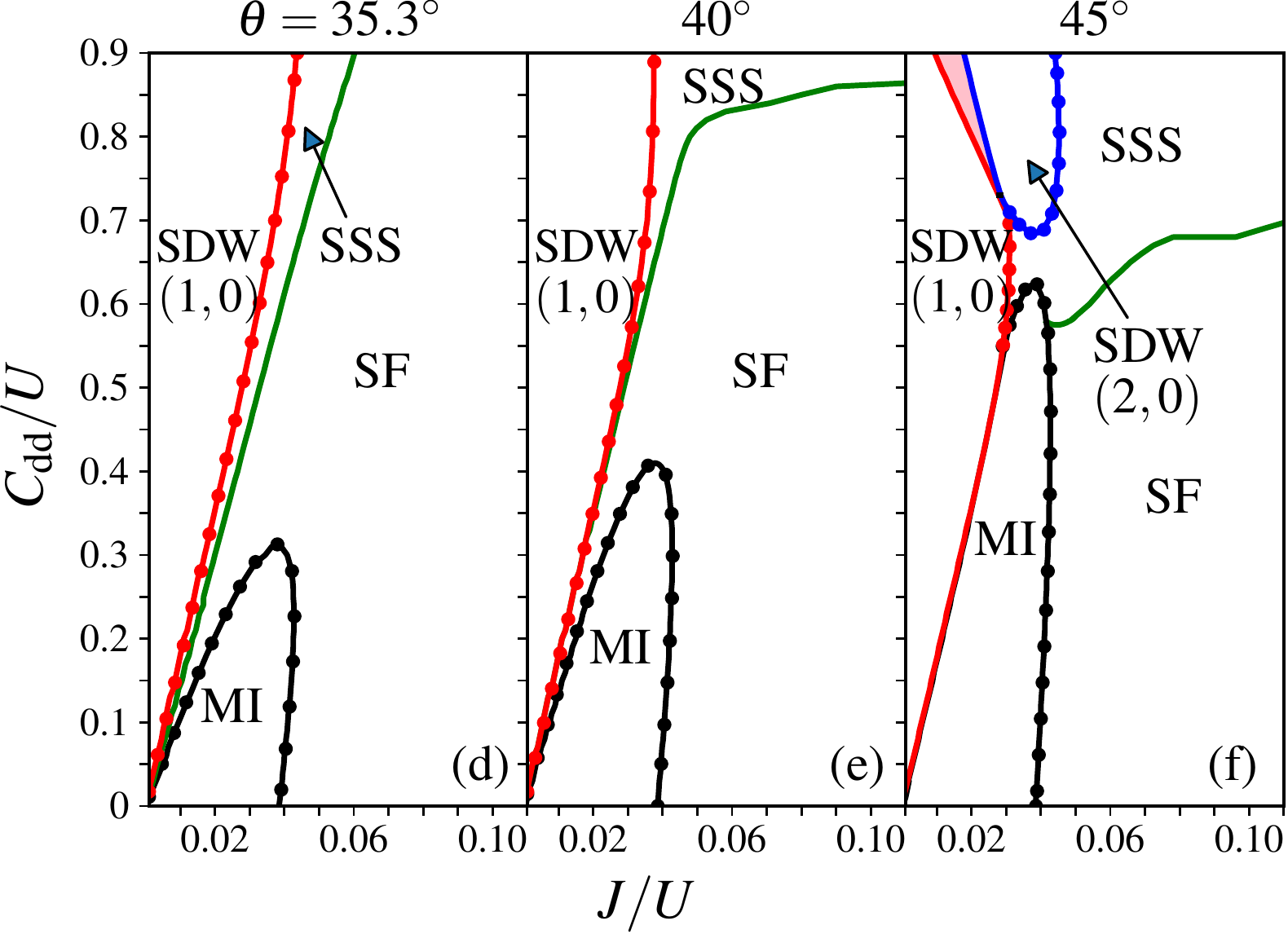}
    \caption{(Color online)
            Shows the phase diagrams in the $J/U-C_{\rm dd}/U$ plane for 
            different values of the tilt angle $\theta$. The phase diagrams are
            obtained for $\mu/J = 15$. The solid phase boundaries are
            obtained from the self-consistent numerical diagonalization of
            mean-field Hamiltonian of the system. Whereas, filled circles
            mark the phase boundaries between an incompressible and a
            compressible phase of the system, which are obtained analytically 
            considering perturbation analysis of the mean-field decoupling 
            theory. In Fig.~(c), the parameter regions for emulsion SS and 
            emulsion DW (1,0) phases are shaded by silver and gold colors 
            respectively. In these emulsion phases, the system simultaneously 
            exhibits both the orders, checkerboard and striped. The parameter 
            region shaded by pink color in Fig~(f) is for the emulsion phase of
            SDW (1,0) and SDW (2,0) phases.
            }
    \label{ph_diags_uos_vdp}
 \end{figure}

 The $J/U$-$C_{\rm dd}/U$ phase diagrams for different values of $\theta$ are 
shown in Fig.~(\ref{ph_diags_uos_vdp}). In the figure, the solid lines 
correspond phase boundaries obtained from the Gutzwiller mean-field theory. 
The filled circles mark the phase boundaries between an incompressible and a 
compressible phase, which are calculated from the mean-field decoupling theory. 
From the figure, it is evident that the mean-field decoupling theory, when 
applicable, gives results which are in good agreement with the Gutzwiller 
mean-field theory. For the parameters considered we obtain MI phase with unit 
filling. The MI-SF phase boundary is obtained by solving 
Eq.~(\ref{mi_sf_boundary}) 
with $n_0 = 1$. The SSS-SDW phase boundaries are calculated by solving 
Eq.~(\ref{sdw_sss_boundary}) with $n_{A} = 1$ and $n_{B} = 0$ for the 
SDW (1,0)-SSS boundary, and  $n_{A} = 2$ and $n_{B} = 0$ for the 
SDW (2,0)-SSS boundary. Similarly, the CBSS-CBDW phase boundaries are calculated
by solving Eq.~(\ref{cbdw_cbss_boundary}) with $n_{A} = 1$ and $n_{B} = 0$ for 
the CBDW (1,0)-CBSS boundary, and $n_{A} = 2$ and $n_{B} = 0$ for the 
CBDW (2,0)-CBSS boundary. 


\subsubsection{$\theta=0^{\circ}$,$15^{\circ}$, and $30^{\circ}$ }
 The phase diagrams for $\theta = 0^{\circ}$, $15^{\circ}$ and $30^{\circ}$ 
are shown in Fig.~(\ref{ph_diags_uos_vdp})(a) to~(\ref{ph_diags_uos_vdp})(c).
These are representative cases for tilt angle lower than the magic angle, that
is, $\theta < 35.3^{\circ}$. For these $\theta$,  $U_{\rm dd}$ is repulsive 
along both $x$ and $y$-directions. The interaction is isotropic when 
$\theta = 0^{\circ}$, and along $y$-axis interaction strength decreases with 
the increase of $\theta$. For lower values of $C_{\rm dd}/U$ the system is 
in DW or SF or MI phase for all values of $\theta$. Out of these, the MI and 
SF phases do not have diagonal order. But, for higher values of $C_{\rm dd}/U$ 
the system favors phases with diagonal order. And, we also get CBSS phase in 
which the system exhibits ODLRO in addition to the diagonal order. In addition,
there are domains in the phase diagram where CBDW phases with different filling 
exist. 

 In the DW phases ODLRO is absent and the system has only diagonal order. 
By comparing the phase diagrams shown in Fig.~(\ref{ph_diags_uos_vdp})(a) 
to~(\ref{ph_diags_uos_vdp})(c), we can infer that the domain with checkerboard 
order diminishes with the increase in $\theta$. This is due to the decrease
in $U_{\rm dd}$, which increases the anisotropy of the dipolar interaction
and checkerboard order becomes energetically unfavourable. 
At $\theta = 30^{\circ}$, Fig.~(\ref{ph_diags_uos_vdp})(c), we get metastable 
{\em emulsion} SS and DW phases. The parameter domains of these phases are 
shaded by the silver and gold colors respectively. In the emulsion phase, the 
checkerboard and striped orders coexist. The emergence of the emulsion phase 
at this tilt angle, implies that $U_{\rm dd}$ is weak and cannot support 
checkerboard order. The system has entered the parameter domain where the 
striped order has lower energy. Indeed, at lower $\theta$ we obtain phases with
striped order. In addition, an important aspect of the phase diagram at 
$\theta = 30^{\circ}$ is the absence of the DW (2,0) phase. It is also to be 
highlighted that, for this $\theta$, the presence of the emulsion phase renders
the mean-field decoupled theory inapplicable to identify phase boundaries 
between incompressible and compressible phases with diagonal order. This is due
to the lack of a well defined unperturbed ground state for the emulsion phase.
However, the presence of the emulsion phase can be identified as the domains 
where Eq.~(\ref{sdw_sss_boundary}) and Eq.~(\ref{cbdw_cbss_boundary}) indicate 
simultaneous presence of striped and checkerboard order in the DW (1,0) phase. 
This overlap region is indicated by the violet filled circles and coincides
with the numerical phase boundary between emulsion SS and emulsion DW (1,0) 
phase. But, this is to be contrasted with the Gutzwiller mean-field results, 
since within this region we obtain a narrow region of CBDW (1,0) phase 
surrounded by the regions of emulsion DW (1,0) phase. It is to be mentioned 
here that the phase diagram for $\theta=0^{\circ}$, shown in 
Fig.~(\ref{ph_diags_uos_vdp})(a), are consistent with the results reported
in our previous work~\cite{suthar_19}. In our previous work, we had explored
the phase diagram of the extended BHM model in the $J/U$-$\mu/U$ plane. And,
thus, parts of the phase diagram for specific values of $C_{\rm dd}/U$ and 
$\mu/J$ in Fig.~(\ref{ph_diags_uos_vdp})(a) corresponds to horizontal cuts of 
the phase diagram reported in ref.~\cite{suthar_19}.


\subsubsection{$\theta=35.3^{\circ}$ and $40^{\circ}$}
  At the magic angle, that is, $\theta = \theta_{\rm M} \approx 35.3^{\circ}$, 
as mentioned earlier, the dipolar interaction along $y$-axis vanishes. But, the
interaction along $x$-axis remains positive and unchanged. Energetically, this 
favours striped order for the phases with diagonal order. And, as shown in 
Fig.~(\ref{ph_diags_uos_vdp})(d), the phase diagram supports SSS and SDW 
phases. For 
$\theta > \theta_{\rm M}$, the dipolar interaction along $y$-axis is 
attractive. This further enhances the striped phases, and this is discernible
from the phase diagram at $\theta = 40^{\circ}$ shown in 
Fig. \ref{ph_diags_uos_vdp}(e). In this case, the SSS phase extends up to 
$J/U\approx 0.2$ for $C_{\rm dd}/U\approx 0.9$.


\subsubsection{$\theta=45^{\circ}$ }
  At higher $\theta$, new stripe phases emerge in the phase diagram, and as
an example we examine the phase diagram at $\theta=45^{\circ}$. As shown in 
Fig. \ref{ph_diags_uos_vdp}(f), SDW (2,0) phase is present in the system when
$\theta=45^{\circ}$. However, at lower $\theta$, the stronger
attractive interaction along $y$-axis results in the instability of the system
and ultimately leads to density collapse.  The phase diagram 
at $\theta=45^{\circ}$ shows two distinct signatures of the onset of the 
instability. First, the mixing of different phases SDW (1,0) and SDW (2,0) in 
the domain shaded by pink color. And, second, the merging of different 
phases MI, SF, SSS and SDW. In contrast, at lower $\theta$ the
incompressible phases are separated by an intervening compressible phase.
It must be mentioned here that, merging of incompressible phases is also 
discussed in previous works on 2D BHM with three-body attractive 
interaction ~\cite{naini_12,singh_18}. The presence of the emulsion phase 
indicates that the phase transition between SDW (1,0) and SDW (2,0) phases is 
not second-order. A detail analysis is essential to understand 
whether the phase transition is first order or a micro-emulsion phase 
intervenes the phases~\cite{spivak_04}.

 In the phase diagram there is a triple point of MI, SDW (1,0) and SSS phases 
at approximately $(0.027,0.5)$. Starting from the triple point there is a sharp 
phase boundary between the MI phase with unit filling and the SDW (1,0) phase 
in the range $0.38\leqslant C_{\rm dd}/U\leqslant 0.50$ and 
$0.021\apprle J/U\apprle 0.027$. This phase boundary can either 
be a first-order phase transition, or a thin region of metastable emulsion 
of the two phases could possibly exist which is not detectable with the
present method. However, for $J/U < 0.021$ and $C_{\rm dd}/U<0.38$, we do 
obtain a very narrow region of the emulsion phase separating these two phases.
\begin{figure}[ht]
    \includegraphics[height = 6.3cm]{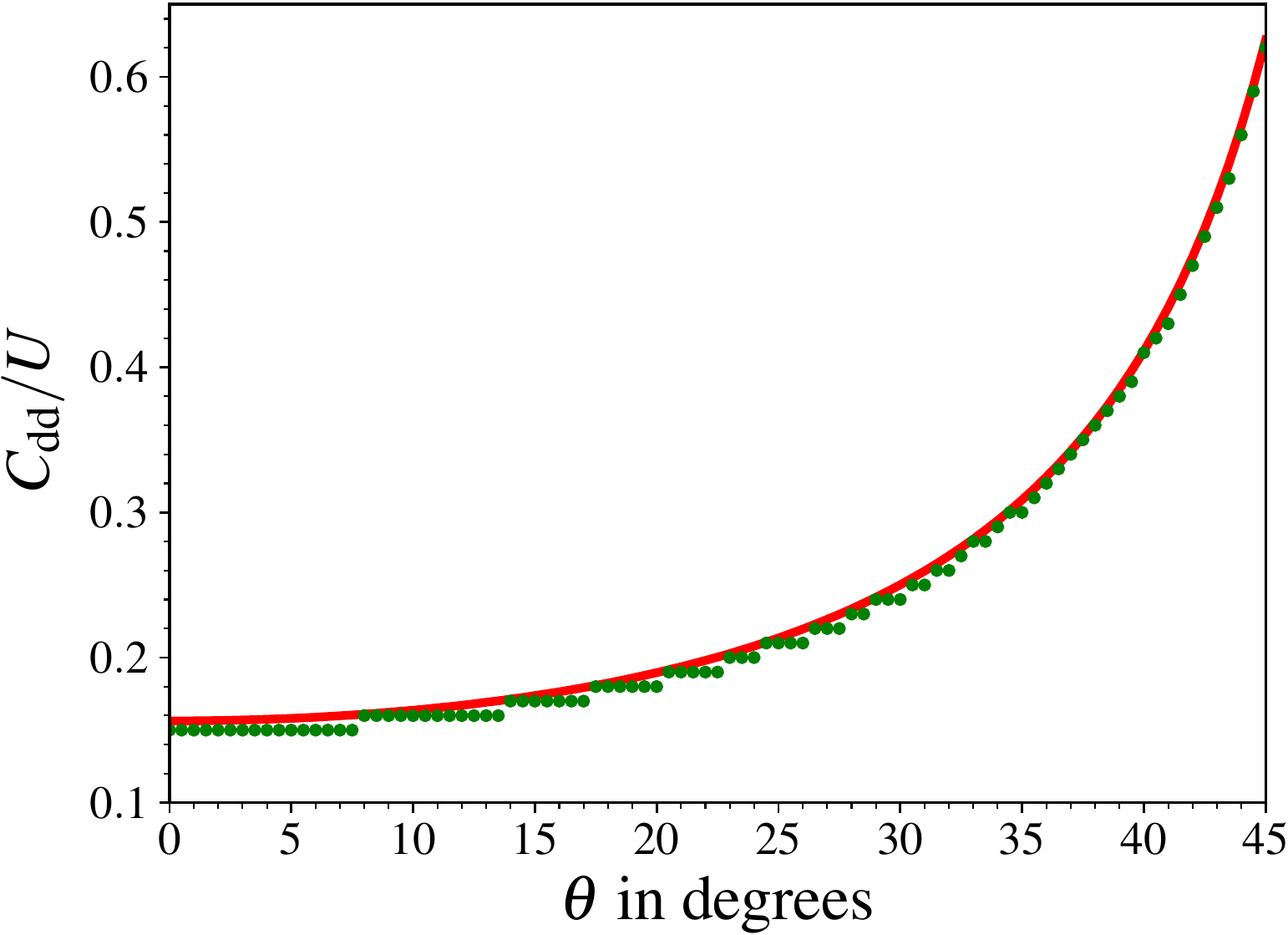}
    \caption{(Color online)
            Shows the $C_{\rm dd}/U$ value of the tip of the MI lobe as the 
            tilt angle $\theta$ is changed. Green filled circles are obtained 
            from Gutzwiller mean-field theory and solid red line is obtained 
            from Eq.~(\ref{mi_sf_boundary}). 
            }
    \label{ml_tip_tilt}
\end{figure}


\subsubsection{MI lobe enhancement }

 One feature of the MI lobe discernible from the phase diagrams in 
Fig.~(\ref{ph_diags_uos_vdp}) is its enhancement along the $C_{\rm dd}/U$-axis
with increasing $\theta$. To illustrate this, the $\theta$ dependence of the MI
lobe tip, in terms of $C_{\rm dd}/U$, is shown in Fig.~(\ref{ml_tip_tilt}). To 
analyze this consider the Eq.~(\ref{mi_sf_boundary}) which defines the MI-SF 
boundary in the mean-field decoupling theory and rewrite it as
\begin{equation}
   \frac{U}{4J} = \left[ \frac{n_0+1}{ n_0 - \tilde{\mu}/U}  
   - \frac{n_0}{(n_0 - 1) - \tilde{\mu}/U} \right]. 
\end{equation}
In absence of the dipolar interaction ($C_{\rm dd}=0$) $\tilde{\mu}=\mu$ and
we obtain the MI-SF boundary of the BHM. However, the dipolar interaction
reduces the effective chemical potential to 
$\tilde{\mu} = \mu -C_{\rm dd}(2 - 3 \sin ^2\theta)$. At $\theta = 0^{\circ}$,
$\tilde{\mu}$ has the smallest value 
$\tilde{\mu}_{\rm min} = \mu -2C_{\rm dd}$ and this can be 
considered as the value of $\tilde{\mu}$ to define the MI-SF boundary. 
But, when $\theta>0^{\circ}$ the prefactor $(2 - 3 \sin ^2\theta)$ decreases
and hence, to maintain the same value of $\tilde{\mu}$ the strength of the 
dipolar interaction $C_{\rm dd}$ has to increase. Thus, there is an 
enhancement of the MI lobe along the $C_{\rm dd}/U$-axis. As the 
degree of enhancement depends on the prefactor with $\sin^2\theta$, the
trend noticeable in Fig.~(\ref{ml_tip_tilt}) is indicative of this dependence.
This is consistent with the experimental finding in ~\cite{baier_16}, where 
onsite repulsive dipolar interaction is observed to favour the MI phase due 
to stronger pinning of the lattice bosons.


\subsection{Phase diagrams in $J/U-\theta$ plane}
\begin{figure}[ht]
    \includegraphics[height = 6.3cm]{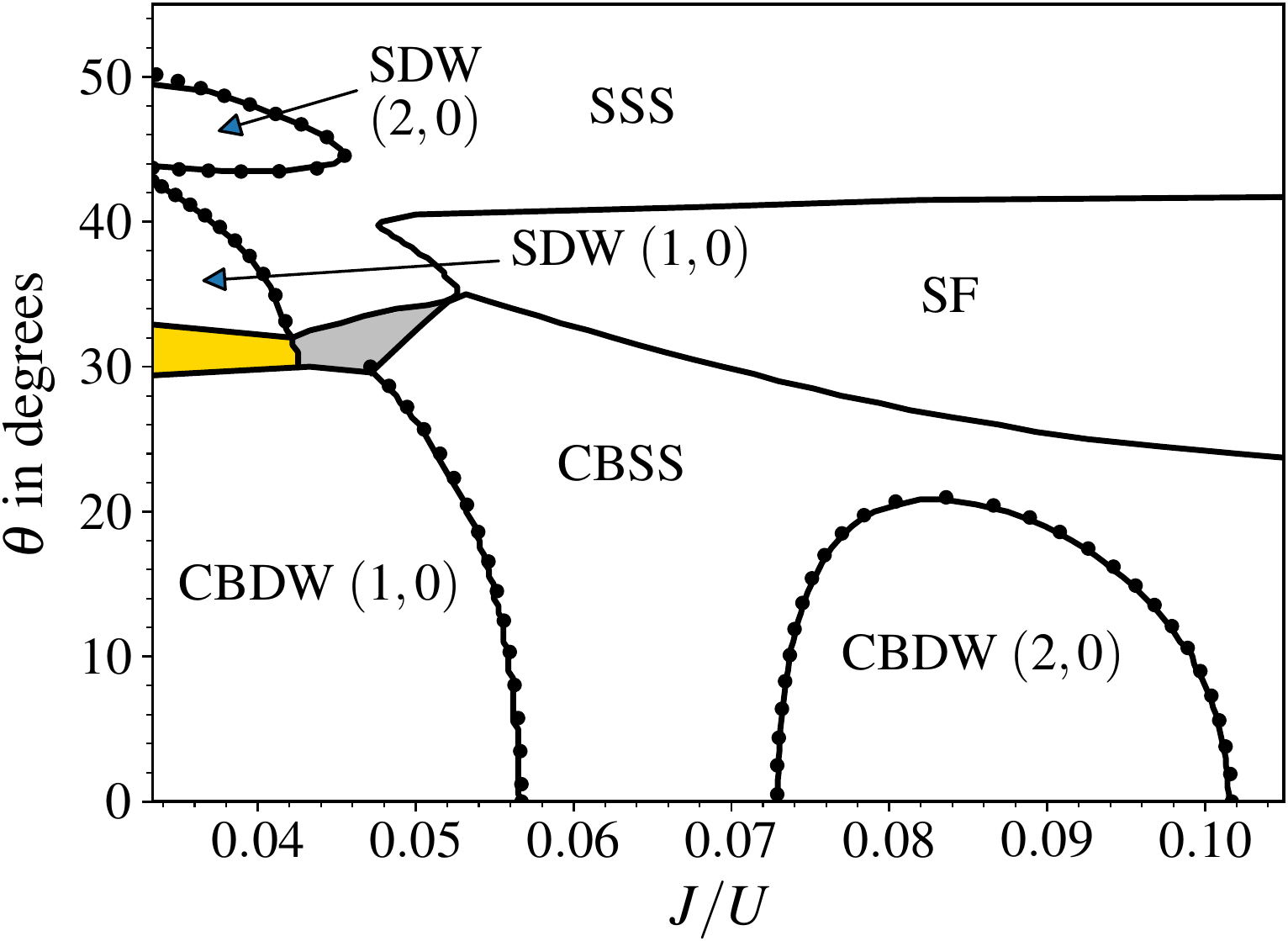}
    \caption{(Color online)
            Shows the phase diagram in the $J/U-\theta$ plane for 
            $C_{\rm dd}/U = 0.8$ and $\mu/J = 15$. The solid phase boundaries 
            are obtained from the numerical computation of the Gutzwiller 
            mean-field theory. The filled circles mark the phase boundaries 
            between an incompressible and a compressible phase which are 
            calculated analytically by performing perturbation analysis of the 
            mean-field decoupling theory. The parameter domains shaded by 
            silver and gold colors are for emulsion SS and emulsion DW (1,0) 
            phases respectively. In these emulsion phases, checkerboard and 
            striped order coexist in the system.
            }
    \label{ph_diag_uos_theta}
 \end{figure}

 From the phase diagrams in Fig.~(\ref{ph_diags_uos_vdp}), it is evident that
the phase structure is richer with stronger dipolar interaction 
(large $C_{\rm dd}/U$). Most importantly, the checkerboard order of the system 
transforms into striped order below a certain value of $\theta$. This is an 
example of structural phase transition. To examine the phases of the system as 
a function of $\theta$ we examine the phase diagram in the $J/U-\theta$ plane
for fixed values of $C_{\rm dd}/U$ and $\mu/J$. And, as an example the phase 
diagram for the case of $C_{\rm dd}/U=0.8$ and $\mu/J=15$ is shown in 
Fig~(\ref{ph_diag_uos_theta}). 
Consistent with the phase diagrams in Fig.~(\ref{ph_diags_uos_vdp}), 
checkerboard and striped orders are preferred for $\theta\apprle25^{\circ}$ 
and $\theta\apprge35^{\circ}$, respectively. 
For $25^{\circ}\apprle\theta\apprle35^{\circ}$ emulsion phase is the 
preferred one in the strongly interacting domain. However, in the weakly 
interacting domain, $J/U\apprge0.053$, SF phase is the intervening phase 
between the checkerboard and striped supersolids. These are in good agreement 
with the previous findings on phase transition between CBDW (1,0) to SDW (1,0) 
in the hardcore limit of the model~\cite{zhang_15}. The intervening emulsion 
and SF phases implies that there is no sharp phase transition between the two 
structured phases. And, also it cannot be a second order phase transition
in the strongly interacting domain, $J/U\apprle0.053$. In this domain, the 
phase transition can either be first order or ``Spivak-Kivelson" type phase 
transition in which a micro-emulsion phase intervenes between two ordered 
phases~\cite{spivak_04}. Considering that the checkerboard order disappears at 
$\theta$ smaller than the magic angle, implies that it is a delicate phase. It 
is unstable against large anisotropy of the interaction potential. 

An important observation, manifest in Fig.~(\ref{ph_diag_uos_theta}), is
the parameter domain of the CBDW (2,0) and SDW (2,0) phases. The former occurs
in the domain of large $J/U$ and small $\theta$. The later, on the other hand,
occurs in the domain with small $J/U$  and large $\theta$. This is, however,
due to the choice of $C_{\rm dd}/U$ and $\mu/J$. For a different choice of 
these two parameters, there could be an intervening emulsion phase for the 
transition between these two structured phases.


\section{Conclusions}\label{conclusions}
 In conclusion, we have explored the rich phase structure of soft core 
dipolar bosons in a 2D optical lattices as a function of tilt angle $\theta$. 
The key point is that the variation of $\theta$ modifies the anisotropy of 
the dipolar interaction in the plane of the 2D lattice. And, this leads
to the formation of two types of quantum phases with different diagonal orders:
checkerboard and striped. Our results indicate that the quantum phase 
transition between these orders, namely, the checkerboard and stripe orders, 
occurs through an intervening emulsion phase. The striped order phases, both 
density wave and supersolid phases, are preferred at high values of $\theta$ 
when the anisotropy is large. However, above the magic angle 
$\theta_{\rm M}\approx 35.3^{\circ}$, as the interaction along $y$-axis turns 
negative, a density instability manifest in the system.

\begin{acknowledgments}
The results presented in the paper are based on the computations using 
Vikram-100, the 100TFLOP HPC Cluster at Physical Research Laboratory, 
Ahmedabad, India. We thank Rashi Sachdeva and S. A. Silotri for valuable 
discussions. RN acknowledges the funding from the Indo-French 
Centre for the Promotion of Advanced Research and UKIERI-UGC Thematic 
Partnership No. IND/CONT/G/16-17/ 73 UKIERI- UGC project. 
KS gratefully acknowledges the support of the
National Science Centre, Poland via project 2016/21/B/ST2/01086.
\end{acknowledgments}

\appendix
 \section{Perturbative treatment of SF order parameter}\label{perturb_ana}
We consider the hopping term in the single-site Hamiltonian as the 
perturbation and the interaction terms along with the chemical potential as 
the unperturbed Hamiltonian. Therefore, the energy of the ground state of the 
unperturbed Hamiltonian
\begin{eqnarray}
  E^{0}_{n_{p,q}} &=& \frac{U}{2}n_{p,q}(n_{p,q}-1) 
                    + \frac{C_{\rm dd}}{2}n_{p,q}(n_{p+1,q}+n_{p-1,q})
                    \nonumber\\
                    &&+ \frac{U_{\rm dd}(\theta)}{2}n_{p,q}(n_{p,q+1}+n_{p,q-1})
                    -\mu n_{p,q}. 
 \label{gs_energy}                   
\end{eqnarray}
Then, to first order in SF order parameter, the perturbed ground state can be
written as 
\begin{eqnarray}
  |\psi_{p,q}\rangle = |n\rangle_{p,q} 
                    + \sum_{m\neq n} 
                    \frac{_{p,q}\langle m|\hat{T}_{p,q}|n\rangle_{p,q}}
                    {E^{0}_{n_{p,q}}-E^{0}_{m_{p,q}}} |m\rangle_{p,q}, 
 \label{perb_state}
\end{eqnarray}
where considering the SF order parameter a real number 
\begin{eqnarray}
 \hat{T}_{p,q}  &=& -J(\phi_{p+1,q}+\phi_{p-1,q}+\phi_{p,q+1}+\phi_{p,q-1})
                    (\hat{b}_{p,q} +\hat{b}^{\dagger}_{p,q})
                   \nonumber\\
                &=& -J\overline{\phi}_{p,q}(\hat{b}_{p,q}
                   +\hat{b}^{\dagger}_{p,q}).
 \label{hop_op}
\end{eqnarray}
Therefore, using Eqs.~(\ref{gs_energy})-~(\ref{hop_op}) the ground state can be
calculated as
\begin{eqnarray}
  |\psi_{p,q}\rangle &=& |n\rangle_{p,q} + J\overline{\phi}_{p,q} \left[
              \frac{\sqrt{n_{p,q}+1}}
                   {U n_{p,q} - \tilde{\mu}_{p,q}}|n_{p,q}+1\rangle\right.
              \nonumber\\
              &&\left.
              - \frac{\sqrt{n_{p,q}}}{
                  U (n_{p,q} - 1) - \tilde{\mu}_{p,q}}|n_{p,q}-1\rangle\right], 
\end{eqnarray}
From this state, we obtain the SF order parameter $\phi_{p,q}$ in the form 
mentioned in Eq.~(\ref{phi_condition}).

\section{CBDW-CBSS phase boundary}\label{cbdw_cbss_pb}

To obtain the phase boundaries between the CBDW and CBSS phases from 
Eq.~(\ref{phi_condition}), we consider checkerboard sublattice structure. 
Then, define $\overline{\phi}_{p,q} = 4\varphi_{B}$ and  
$\tilde{\mu}_A = \mu - \left [C_{\rm dd} + U_{\rm dd}(\theta)\right ]n_{B}$ 
for $(p,q)\in A$ sublattice, and $\overline{\phi}_{p,q} = 4\varphi_{A}$, 
$\tilde{\mu}_B = \mu - \left [C_{\rm dd} + U_{\rm dd}(\theta)\right ]n_{A}$ 
for $(p,q)\in B$ sublattice. This leads to two coupled equations 
\begin{subequations}
   \begin{equation}
     \varphi_{A} = 4J\varphi_{B} \left[ \frac{n_{A}+1}{U n_{A} - \tilde{\mu}_A}
     - \frac{n_{A}}{U (n_{A} - 1) -\tilde{\mu}_A} \right],
   \label{cbdw_cbss_phi_a}
   \end{equation}
   \begin{equation}
    \varphi_{B} = 4J\varphi_{A} \left[ \frac{n_{B}+1}{U n_{B} - \tilde{\mu}_B}
    - \frac{n_{B}}{U (n_{B} - 1) - \tilde{\mu}_B} \right].
   \label{cbdw_cbss_phi_b}
   \end{equation}
\end{subequations}
These two equations can be solved simultaneously. In the CBSS phase 
$\{\varphi_{A}, \varphi_{B}\}\to0^{+}$ across the CBDW-CBSS phase boundary. 
Then, the CBDW-CBSS phase boundary is obtained as in 
Eq.~(\ref{cbdw_cbss_boundary}).

 \bibliography{dipolar}{}
\bibliographystyle{apsrev4-1}

\end{document}